\documentclass{article}

\usepackage{PRIMEarxiv}

\usepackage[utf8]{inputenc} 
\usepackage[T1]{fontenc}    
\usepackage{hyperref}       
\usepackage{url}            
\usepackage{booktabs}       
\usepackage{amsfonts}       
\usepackage{nicefrac}       
\usepackage{microtype}      
\usepackage{lipsum}
\usepackage{fancyhdr}       
\usepackage{graphicx}       
\graphicspath{{media/}}     

\title{NP4G : Network Programming for Generalization
}

\author{
  Shoichiro Hara \\
  Nagoya City University \\
  \texttt{s.hara@nsc.nagoya-cu.ac.jp} \\
   \And
  Yuji Watanabe \\
  Nagoya City University \\
  \texttt{yuji@nsc.nagoya-cu.ac.jp} \\
}

\begin{document}
\maketitle

\begin{abstract}
 Automatic programming has been actively studied for a long time by various approaches including genetic programming. 
 In recent years, automatic programming using neural networks such as GPT-3 has been actively studied and is attracting a lot of attention. 
 However, these methods are illogical inference based on experience by enormous learning, and their thinking process is unclear. 
 Even using the method by logical inference with a clear thinking process, the system that automatically generates any programs has not yet been realized. 
 Especially, the inductive inference generalized by logical inference from one example is an important issue that the artificial intelligence can acquire knowledge by itself.
 
 In this study, we propose NP4G: Network Programming for Generalization, which can automatically generate programs by inductive inference. 
Because the proposed method can realize "sequence", "selection", and "iteration" in programming and can satisfy the conditions of the structured program theorem, it is expected that NP4G is a method automatically acquire any programs by inductive inference.
 
 As an example, we automatically construct a bitwise NOT operation program from several training data by generalization using NP4G. 
 Although NP4G only randomly selects and connects nodes, by adjusting the number of nodes and the number of phase of "Phased Learning", we show the bitwise NOT operation programs are acquired in a comparatively short time and at a rate of about 7 in 10 running.
 The source code of NP4G is available on GitHub as a public repository\footnotemark[1].
\end{abstract}

\keywords{Inductive Inference \and Automatic Programming \and Knowledge Acquisition \and Genetic Programming \and Genetic Network Programming}

\footnotetext[1]{https://github.com/Amplil/np4g}
\section{Introduction}
Automatic programming has been actively studied for a long time by various approaches including genetic programming.
In recent years, automatic programming using neural networks such as GPT-3\cite{gpt3} has been actively studied and is attracting a lot of attention.
However, these methods are illogical inference based on experience by enormous learning, and their thinking process is unclear.
Even using the method by logical inference with a clear thinking process, the system that automatically generates any programs has not yet been realized.

In order to realize automatic programming based on this logical inference, it is necessary to extract the structure of some kind of solutions to the problem, so that this field is closely related to the knowledge acquisition of artificial intelligence.
In the study on the knowledge acquisition of the artificial intelligence, there is still a big problem that knowledge including a lot contradiction and/or an exception is hard to formulate.
Especially, the inductive inference generalized by logical inference from one example is an important problem that the artificial intelligence can acquire knowledge by itself.

In this study, we propose NP4G: Network Programming for Generalization, which can automatically generate programs by inductive inference.
Because the proposed method can realize "sequence", "selection", and "iteration" in programming and can satisfy the conditions of the structured program theorem, it is expected that NP4G is a method automatically acquire any programs by inductive inference.
As an example, we automatically construct a bitwise NOT operation program from several training data by generalization using NP4G.
Although NP4G only randomly selects and connects nodes, by adjusting the number of nodes and the number of phase of "Phased Learning", we show the bitwise NOT operation programs are acquired in a comparatively short time and at a rate of about 7 in 10 running.

Section 2 explains automatic programming, inductive inference, and genetic programming as related research.
Section 3 proposes network programming for generalization (NP4G) and explains its basic structure.
In Section 4, as an example of the proposed method, we describe how to acquire a bit NOT operation program, and in Section 5, we show and discuss the verification results.
Section 6 describes the significance of this study and future issues.
The source code of NP4G is available on GitHub as a public repository \footnotemark[1]. 

\section{Related Research}
\subsection{Automatic Programming}
\label{sec:headings}
Automatic programming (AP) is the automation of all or part of the generation of programs, and achieve measurable success as an aid to developers of large systems and small programs \cite{AutomaticProgramming}. 
Among automatic programming, a model that automatically generates programs by logical inference has been researched for a long time.
Logic Theorist, the world's first artificial intelligence program published in 1956, was designed to imitate human logical inference using search trees and heuristics \cite{LogicTheorist}.

As research on automatic programming that has been attracting attention in recent years, there are methods using a large-scale language model based on a neural network, represented by GPT3 \cite{gpt3} of OpenAI.
The methods predict and automatically generate code according to the situation by learning as a language model by deep learning from a huge amount of publicly available code.
The methods have been very successful in applications such as predicting the code that the developer is going to write on the editor and suggesting the continuation of the code \cite{copilot}.
However, the methods do not generate code by logical inference because they are language generation models and are inference based on illogical experience with huge amount of learning.
In this respect, many problems still remain today.

Recently, some researches have been achieved in the generation of automatic programming by logical inference \cite{palsql}, however there remains a problem in terms of acquiring arbitrary programs since the methods are effective only for domain-specific languages such as SQL.

\subsection {Inductive Inference}
In a wide sense, logical inference consists of deductive inference, inductive inference, and analogical inference \cite{math300}.
Representative examples of logical thinking include "inductive thinking", "analogous thinking", "thinking of generalization", and "symbolic thinking" \cite{saito:11}.
Of these, "thinking of generalization" corresponds to inductive inference.
Inductive inference is an inference that derives general rules to explain given data \cite {inductive-reasoning}.
Inductive inference by artificial intelligence has been studied for a long time \cite{CASE1983193}\cite{4767034}, but its scope of application is limited and problems remain in acquiring arbitrary programs.
In addition, since inductive inference by artificial intelligence can be said to be a method of automatically acquiring knowledge, it is closely related to research on knowledge acquisition.
Many knowledge cannot be formulated, and even if formulated, there are contradictions between rules. 
Such problem is called knowledge acquisition problem \cite{KnowledgeAI}\cite{KAIssues}, and it is still not resolved.

From the point of view of inference with generalization, neural network methods are also within the scope of the field because they can pick out common features of individual data.
However, it is not logical because the correlation is obtained based on a huge amount of data.
In addition, the decision process is poorly explained, and this problem is called the black-box problem \cite{BlackBoxProblem}.
And it requires a large amount of data for learning, so it is only an inference that generalizes, but not an inductive inference.

\subsection {Genetic Programming}
Genetic programming (GP) is a method for automatically generating tree-structured programs by using genetic manipulations \cite{Koza1994}.
It is used to solve various problems such as the automatic construction of formulas and the generation of agent action sequences.
GP handles only tree structures, but genetic network programming (GNP) is an extension of GP to networks \cite{gnp}.
GP/GNP regards each node in the network as the minimum unit for simple processing, and automatically changes the way they are combined to build more optimal programs.
Each node can be classified into a decision node, a processing node, and a start node.
GP/GNP can automatically acquire a program represented by a network, but there is no example of its application to inductive inference.

\section {Proposed Method}
\subsection{Basic Concept of NP4G}
\begin{figure}[t]
\begin{center}
\includegraphics[width=80mm]{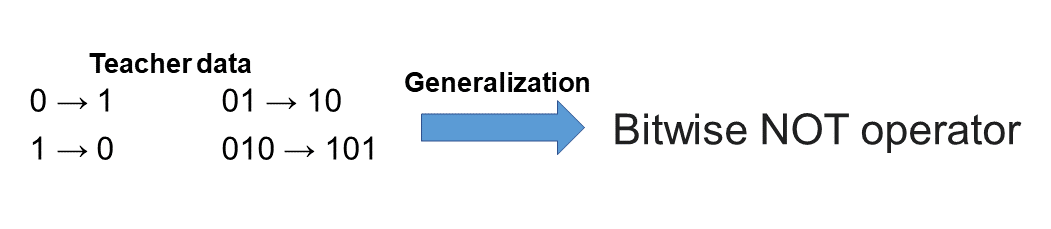}
\end{center}
\caption {Generalization from teacher data}
\label{fig:summary}
\end{figure}

NP4G is a method of performing inductive inference by automatically generating a program represented by a network based on training data.
For example, in Figure \ref{fig:summary}, a bitwise NOT operation program is obtained by generalization from four training data.
The training data is one-input, one-output data, and the search is performed until a program is generated that can obtain inputs and outputs that match the training data.
The generated program is obtained by connecting multiple nodes with simple functions in a network, similar to the concept of GNP.
GNP is network programming using genetic methods, but NP4G is a method that assumes the application of various methods, not limited to the use of genetic methods.

NP4G can be said to be a method of knowledge acquisition by inductive inference in that it acquires one piece of knowledge (program) by generalizing each case (training data).
Therefore, unlike methods using neural networks, the number of required training data is very small, only a few that can grasp the characteristics of the target to be generalized.
In addition, since the constructed program itself is in the form of a network, it is clear what kind of processing is being done internally.

Attempts to automatically generate programs by combining networks have so far been used only in the genetic method of GNP.
However, the proposed method uses network programming as a means of generalization.
Network programming, which is not limited to this genetic method, is expected to be widely applied in new fields in the future as a new automatic programming method.
In the future, similar to GP/GNP, it is expected to be extended to effective algorithms by combining with other methods such as neural networks and reinforcement learning.

\subsection {Structure Theorem}
There is a problem that the method that can make logical inference has a limited application.
However, NP4G has a structure that satisfies the conditions of the structure theorem in order to obtain an arbitrary program.
Structure theorem states that any program can be constructed by combining three types of basic structures: "sequence", "selection", and "iteration" \cite{StructuredProgramming}.
As explained in subsection \ref{sec:struct}, NP4G can achieve "sequence" by executing nodes in a predetermined order from the start node, "selection" by providing decision nodes and processing nodes as preliminarily provided functions, and "iteration" by iterable data. 
So it can be said that the condition of the structure theorem is satisfied.

\subsection{Basic Structure of NP4G}
\label{sec:struct}
NP4G has a directed graph structure, such as Figure \ref{fig:sequence}, in which nodes such as multiple functions and objects are connected in a network.
Nodes are functions (squares in the figure), a start node (a circle marked with "S" in the figure) with input data that is executed at the beginning of the program , a end node (a circle marked with "E" in the figure) that signal the end of the program, and object nodes (circles in the figure) that output pre-stored data without an input link.
The number of links connected as inputs to a node varies depending on the function, but any number of links can be connected as outputs regardless of the function.

\subsubsection {Execution Order of Nodes}
\label{sec:sequence}
\begin{figure*}[t]
\begin{center}
\includegraphics[width=130mm]{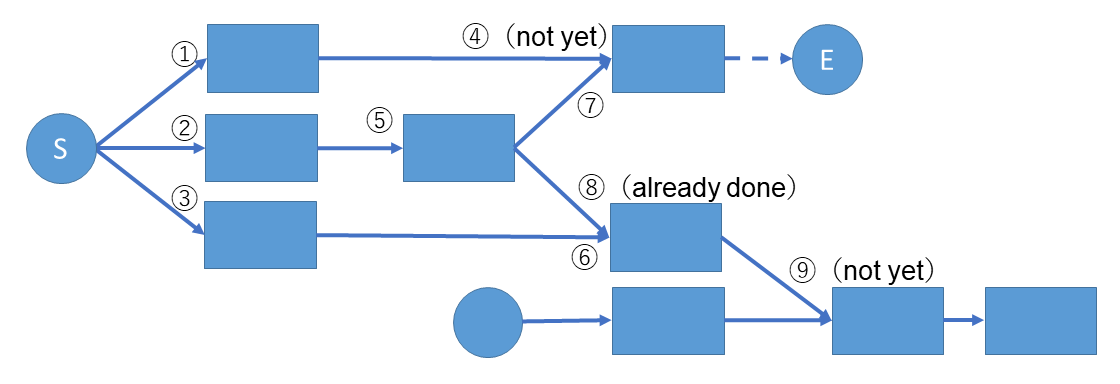}
\end{center}
\caption{Basic Structure of NP4G and Execution Order of Nodes}
\label{fig:sequence}
\end{figure*}

Using an example of a network generated by NP4G shown in Figure \ref{fig:sequence}, we explain the execution order of nodes.
By processing the nodes in order from the start node, the "sequence" of the structure theorem can be realized.

First, the nodes connected to the start node with input data are executed in order from the top, starting with \textcircled{\scriptsize 1}, \textcircled{\scriptsize 2}, \textcircled{\scriptsize 3}.
When all the nodes are executed, the nodes connected to them are also executed in order from the top, \textcircled{\scriptsize 4}, \textcircled{\scriptsize 5}, \textcircled{\scriptsize 6}.
When the network is regarded as a data series, this operation to be executed in order from the top means the operation to arrange the nodes to be executed next as a list and execute them in order from the front.

Here, like \textcircled{\scriptsize 4} in Figure \ref{fig:sequence}, if there is no output from other input nodes, the execution result will be "not yet" and no output will occur.
This means that if there are nodes that do not derive from the start node, such as \textcircled{\scriptsize 9}, they will not be output.
On the other hand, in the case of \textcircled{\scriptsize 6}, \textcircled{\scriptsize 5} was executed immediately before, and all the outputs of the input nodes exist, so they are output.
Next \textcircled{\scriptsize 7}, \textcircled{\scriptsize 8}, and  \textcircled{\scriptsize 9} are executed, but in the case of \textcircled{\scriptsize 8}, this node has already been 
 output in \textcircled{\scriptsize 6}, the execution result will be "already done" and will not be output again.
Each node is executed in this way until there are no more nodes to process next.

Finally, the last output node is responsible for the output of the entire network and is connected to the end node.
In other words, the end nodes are not connected to the network from the beginning, but are determined after the fact from the relation of the execution order of the nodes.

\subsubsection {Iteration}
In order to satisfy the structure theorem, it is necessary to introduce "iteration".
In NP4G, iterative processing is realized not by providing a feedback loop on the network, but by inputting iterable data to a function.
A function given an iterable data as input does the same process for each element of the data.
A function also creates iterable data or converts it back to non-iterable data, and the process is repeated as long as iterable data is used as the function input.
By doing so, iteration can be achieved without creating a program that gets stuck in an infinite loop and never terminates.
In addition, even when randomly combining nodes and automatically generating a program, the simplicity of the node execution order makes it less likely that errors occur.

\subsubsection {Automatically Defined Functions (ADFs)}
GP uses Automatically Defined Functions (ADFs) because they can be expected to be faster as programs evolve \cite{adfs}.
ADFs make it possible to create more advanced programs in a short period of time by reusing networks that have already been created.
NP4G combines ADFs with phased learning described in \ref{sec:PL}, registering networks as ADFs at each phase and allowing the networks to be reused in the next phase.

\subsubsection {Phased Learning}
\label{sec:PL}
Phased learning is a method of learning even a complicated program without difficulty by dividing learning into each phase.
Phased learning is a method mainly used in reinforcement learning, and it has the effect of preventing learning time from becoming enormous by lowering the degree of freedom that learning can take.

NP4G starts by reducing the number of training data to be learned and generating a network that performs simple processing.
Then, the generated networks are reused as ADFs when constructing the next network.
By doing so, we can expect to learn in a shorter time than when obtaining a complicated network from the beginning.

\section {Acquisition of a Bitwise NOT Operation Program}
Using NP4G, we consider the case of automatically constructing a bitwise NOT operation program, and verify it by actually executing the NP4G program.
In this study, we use the programming language Python (Python 3.7.12) to create a NP4G program.
We use Google's Colaboratory as the execution environment.
In addition, list-type objects (hereafter referred to as lists) are used in Python programs as iterable data for realizing iterative processing.
In NP4G, all data stored in training data, start nodes, and object nodes are character strings.

\subsection {Preliminarily Provided Functions}
\begin{figure*}[t]
\begin{center}
\includegraphics[width=110mm]{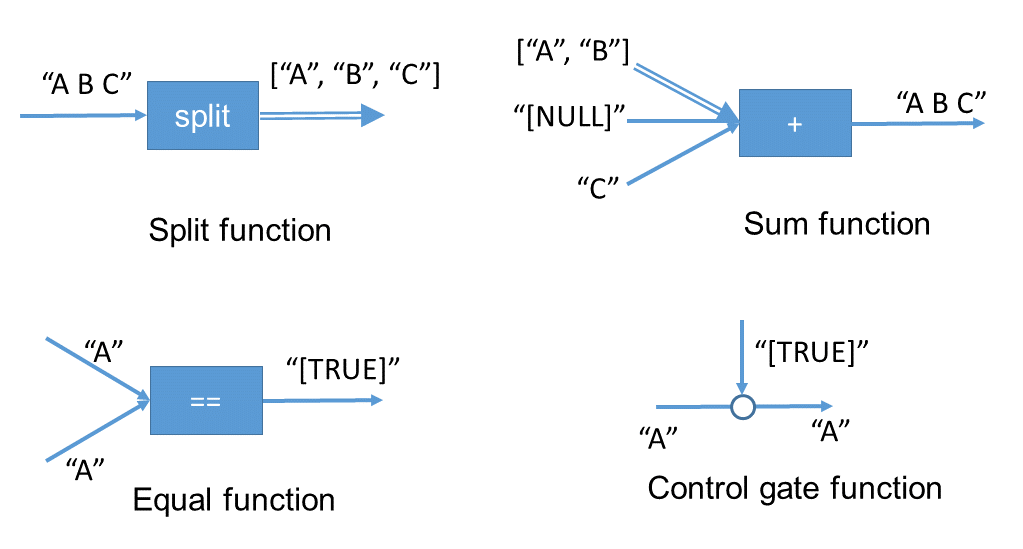}
\end{center}
\caption {Function to be given beforehand}
\label{fig:func}
\end{figure*}

As a function to construct the network, four nodes are given in advance: split function, sum function, equal function, and control gate function, as shown in Figure \ref{fig:func}.
The contents of each function are explained below.

\subsubsection{Split Function}
Split function separates strings with spaces as delimiters and write them as a list.
In figures, "split" is enclosed in a square.

\subsubsection{Sum Function}
Sum function smoothes multiple inputs, such as list-type inputs and other character strings, and combines them with a space between them to make one character string.
If the input string is "[NULL]", then "[NULL]" is not concatenated.
If the output string is "" (empty), then "[NULL]" is outputted.
In figures, "+" is enclosed in a square.

\subsubsection {Equal Function}
Equal function outputs "[TRUE]" when the values of the two inputs match.
Otherwise, it outputs "[FALSE]".
It plays the role of decision node in GP/GNP.
In figures, "==" is enclosed in a square.

\subsubsection {Control Gate Function}
If one of the two inputs is "[TRUE]", control gate function passes the value of the other input, otherwise outputs "[NULL]".
It plays the role of a processing node in GP/GNP.
In figures, it is indicated by white circles.

\subsection {Example of a Bitwise NOT Operation Program}
\begin{figure*}[t]
    \begin{center}
        \includegraphics[width=150mm]{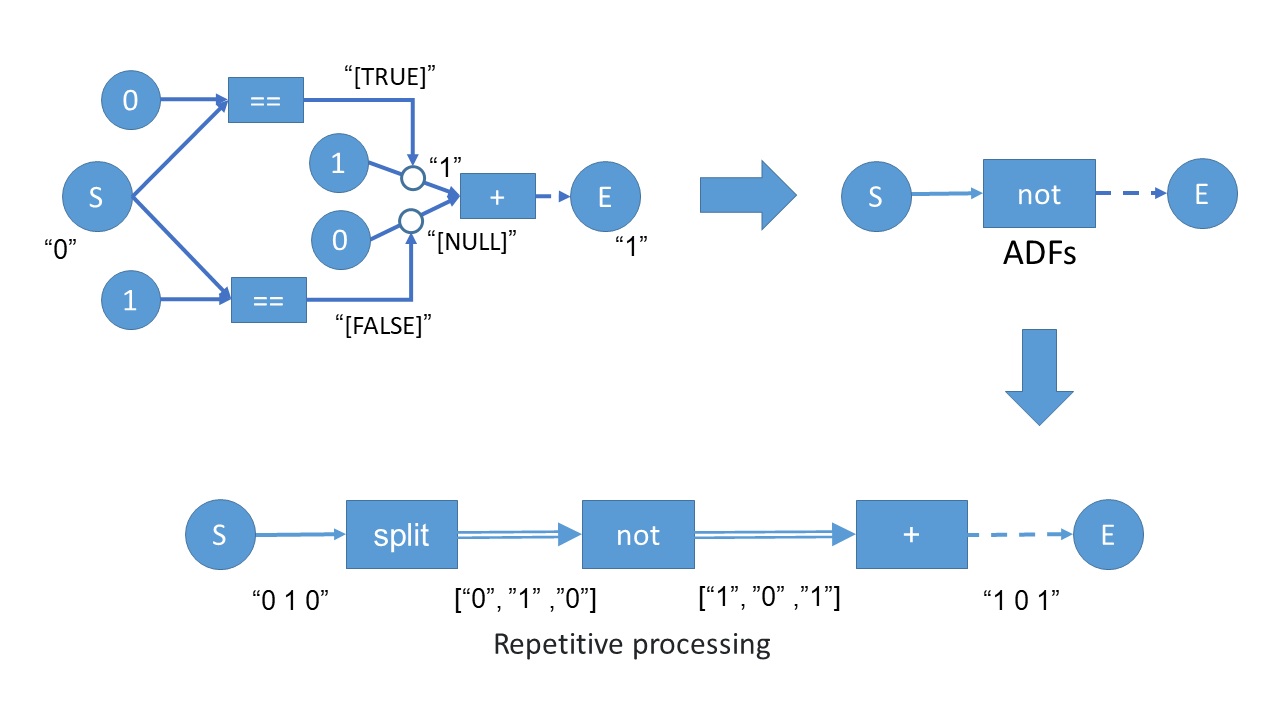}
    \end{center}
    \caption{Implementation of a bit NOT operation program by NP4G}
    \label{fig:bitwise_not}
\end{figure*}

Figure \ref{fig:bitwise_not} shows an example of a bitwise NOT operation program realized by using these four functions.
As the first phase, NP4G constructs the network in the upper left of the figure for 1 bit input.
The equal function whose input is an object node containing "0" and a start node containing input "0" outputs "[TRUE]".
The control gate function with "[TRUE]" as input outputs the value of the object node storing "1" as it is.
In this way, the "selection" of the structure theorem is realized by the equal function and the control gate function.
If the input is "1", another equal function and a control gate function that output "[FALSE]" and "[NULL]" for input "0" output "[TRUE]" and "0" respectively.
In this way, a 1 bit logical NOT is constructed as a network.

Next, as shown in the upper right of the figure, this network is made into a form that can be reused for building the next network as ADFs.
In the figure, "not" is indicated by enclosing it in a square.
Then, in the next phase where multiple bits are input, NP4G constructs a bitwise NOT operation program by using the iterative processing of iterable data (list) and the ADFs "not".
Multi-bit input strings are made into a list by the split function, the ADFs "not" are repeatedly applied to each element, and the disjoint strings are combined by the sum function to realize multi-bit logical not.
By phased learning that builds a network phase by phase in this way, the target program is realized.

\subsection {Verification Method}
We verify whether a bitwise NOT operation program can be automatically acquired by NP4G simply by randomly selecting and connecting nodes for several training data.

\subsubsection {Training Data and Random Search}
The training data used are only four shown in Table \ref{tbl:TeacherData}, and the training data used at each phase changes depending on the method of phased learning.
Each input and output of the training data is written in parentheses such as (input, output).

\begin{table}[htbp]
\centering
\caption {All training data to use in this study}
\label{tbl:TeacherData}
\begin{tabular}{l}
\hline
(input, output) \\
\hline \hline
( "0" , "1" ) \\
( "1" , "0" ) \\
( "0 0" , "1 1" ) \\
("0 1 0" "1 0 1") \\
\hline
\end{tabular}
\end{table}

As a bitwise NOT operation program is automatically generated by NP4G, a network that matches these training data at each phase is searched by randomly combining nodes.
The random algorithm first places a start node and an end node, selects nodes at random for the determined number of nodes, and then connects them with other nodes at random for the number of inputs of the function.

\subsubsection {Method of the Phased Learning}
\label{sec:PLhow}
In this study, we prepare the phases used in phased learning from phase 1 to phase 5, and examine the effect of phased learning by changing the combination of each phase.
The training data for the five phases used in phased learning are as follows.

\begin{itemize}
\item Phase 1: ("0", "1")
\item Phase 2: ("1", "0")
\item Phase 3: ("0", "1"), ("1", "0")
\item Phase 4: ("1", "0"), ("0 0", "1 1")
\item Phase 5: ("1", "0"), ("0 0", "1 1") ("0 1 0", "1 0 1")
\end{itemize}

Then each of these phases is compared by changing the number of phases like follows.
\begin{itemize}
\item 3 phases:
Phase 2, 3, 4
\item 4 phases:
Phase 1, 2, 3, 4
\item 5 phases:
Phase 1, 2, 3, 4, 5
\end{itemize}
\noindent

The networks obtained at each phase can be reused in subsequent phases by adding them to a list as ADFs using Python lambda expressions.
A network that matches the training data at each phase is searched by randomly combining preliminarily provided functions, object nodes, and ADFs.
Note that the object node also needs to be prepared in advance in the same way as preliminarily provided functions.
In the case of object nodes, both input and output character strings of training data used in learning at each phase are prepared as object nodes.

\subsubsection {Verification}
In the verification, the NP4G program is executed 10 times, and the average values of the generation results and execution time are obtained.
The generated result is "success" if the network obtained by searching is a generalized network that can obtain the expected output, and "failure" otherwise. If it is not generated within the time limit for one program execution, it is considered as "exceeded".
To check whether the output is the expected one, enter all binary strings from 1 to 5 digits ("0" to "1 1 1 1 1") as verification data.
Also, the time limit for one execution is 3 hours (10,800 seconds).

\section {Results and Discussion}
Table \ref{tbl:result1} shows the execution results of the NP4G program when the number of phases and the number of nodes are changed.
The execution results are displayed with the numbers of "success", "failure", and "exceeded".
Table \ref{tbl:result2} shows the average, maximum, and minimum execution times (s) when the number of phases and the number of nodes are changed.
From Table \ref{tbl:result1} and Table \ref{tbl:result2}, when there are 4 phases and 20 nodes, the number of successes is as high as 7 and the average time is relatively short as 1,105s.
In the case of 4 phases and 15 nodes, the number of successes is 6, but the average time is the shortest at 334s.
These are the best results.

As shown in Table \ref{tbl:result1}, there is only one failed network when there are 4 phases and 10 nodes.
Also, from the results when the number of nodes is 5, all of them exceed the time except for one successful case of 5 phases.
This is probably because the number of nodes is too small to generate a bitwise NOT operation program with 5 nodes.
On the other hand, when the number of nodes is large, such as 20, the number of time overruns tends to increase, which is thought to be due to the time required to generate the network as the number of nodes increases.
Next, looking at the average time for each number of phases in Table  \ref{tbl:result2}, we can see that the average time for 3 phases tends to be long.
The reason for this is thought to be that the number of phases to be taken is small in 3 phases, and it becomes necessary to create a complicated network in one phase.
Also, the reason why the average time tends to be longer for 5 phases than for 4 phases is that if the number of phases is increased too much, it will take extra time.

\begin{figure*}[t]
\begin{center}
\includegraphics[width=150mm]{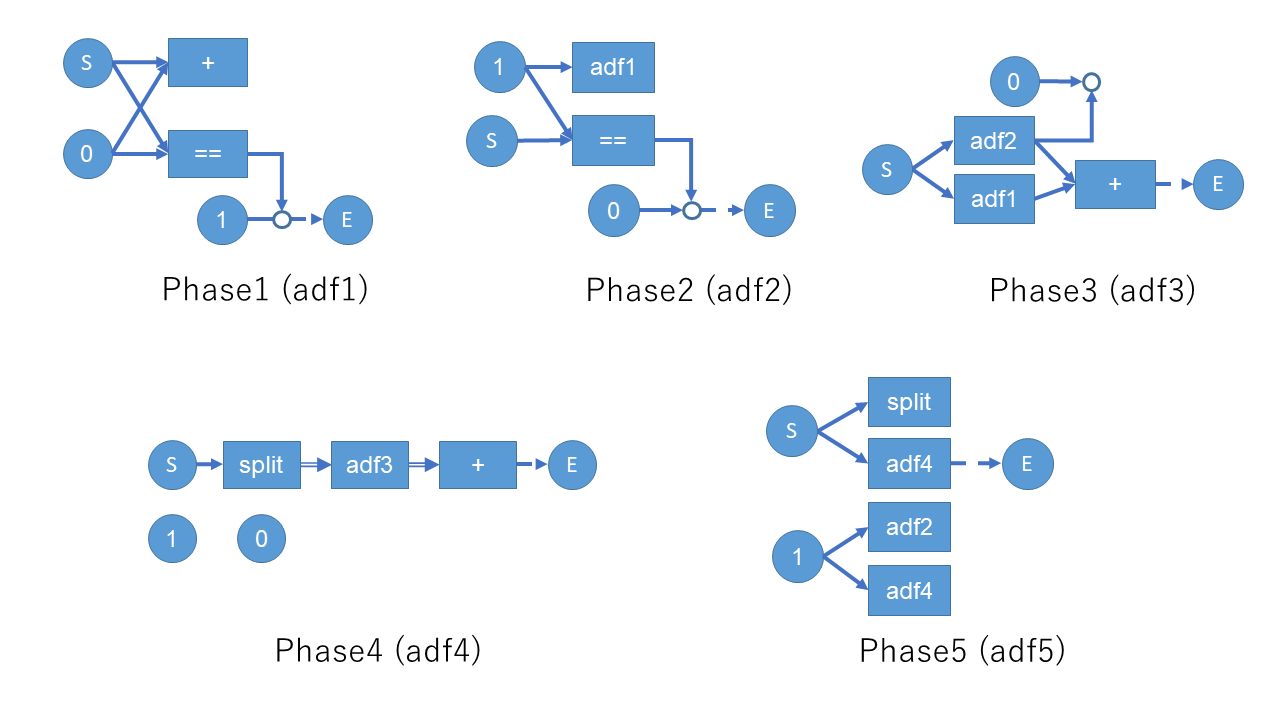}
\end{center}
\caption {Network when the expected program is obtained by automatic generation at 5 phases with 5 nodes}
\label{fig:out_net_p5n5}
\end{figure*}

Figure \ref{fig:out_net_p5n5} shows a 5-nodes network with 5 phases that is the only successful automatic generation.
It is shown that the expected network can be obtained even if the number of nodes is small like this network.
After automatically generating a network with an output of 1 for an input of 0 in Phase 1, it is used as adf1 in Phase 2, and then the network created in Phase 2 is used as adf2 in Phase 3.
In Phase 4, we can see that adf3 is used to create a network similar to the example implementation of a bitwise NOT operation program in Figure \ref{fig:bitwise_not}.
In the case of the network, we confirm that a bitwise NOT operation program is obtained in Phase 4, and the expected network is obtained even in the middle of phased learning.

\begin{figure*}[t]
\begin{center}
\includegraphics[width=150mm]{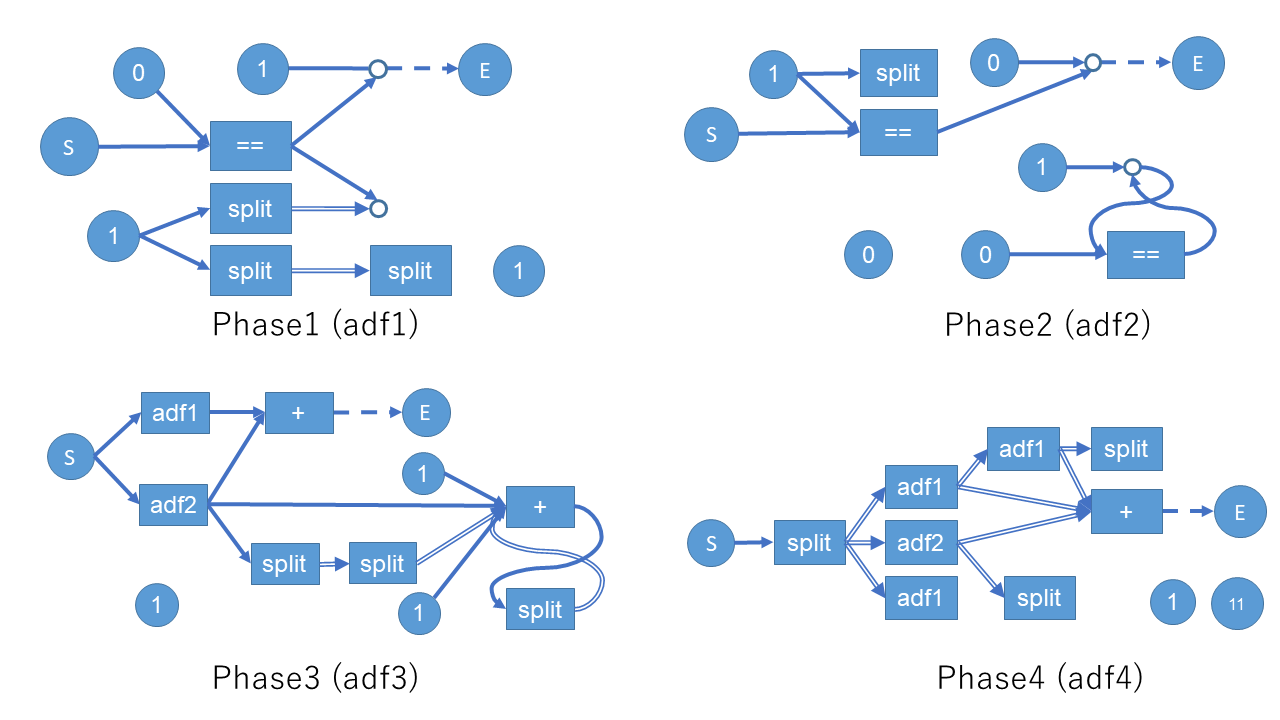}
\end{center}
\caption {A network obtained by automatic generation with 10 nodes in 4 phases, where the training data used for learning matches the input and output, but a program different from expectations is obtained.}
\label{fig:out_net_p4n10}
\end{figure*}

Next, Figure \ref{fig:out_net_p4n10} shows the network when automatic generation with 10 nodes in 4 phases fails.
In this way, because the network becomes complicated, it is thought that although there are outputs that match all the training data, there are outputs that do not match the verification data.
In the case of this network, an input of "0 1 0" results in "0 1 1" and an output that is not "1 0 1".
Even in this case, it is considered that the expected network can be obtained by increasing the number of phases and learning with training data including ("0 1 0", "1 0 1").
Also, from Phase 2 and Phase 4 in Figure \ref{fig:out_net_p4n10}, we confirm that the ADFs adf1 and adf3 generated in the previous phase are not always used because the nodes are randomly selected when constructing the network.
As we can see from the networks obtained in the actual verification of Figure \ref{fig:out_net_p5n5} and Figure \ref{fig:out_net_p4n10}, NP4G has a clear thinking process, unlike methods using neural networks.

Table \ref{tbl:result3} shows the average, maximum, and minimum generation times (s) at each phase.
This table is compiled from all successful results regardless of the number of nodes and phases.
From this table, it can be seen that the generation times in Phase 1 and 2 are shorter than the other phases.
This is because both Phase 1 and 2 are learning with only one set of training data, so it is easy to find a network that matches the training data even if the network is generated randomly.
Next, it can be seen that the average generation time in Phase 3 is the longest at 2585.24s, followed by Phase 4, and then Phase 5.
We consider these factors as follows:
First, Phase 3 is the learning of training data ("0" "1") and ("1", "0"), and for the first time with one network, it is necessary to search for a network that satisfies two conditions.
Therefore, it is considered that it is necessary to generate a complicated network.
In Phase 4, a complicated network is not required, but the generation time is considered to be longer because it is necessary to generate a network that required iteration processing for the first time in Phase 4.
In the case of Phase 5 after getting the iterative process, even if the number of digits increases, the same network can be used, so it is thought that the generation time can be shortened.

\begin{table}[htbp]
\centering
\caption {Execution result (success / failure / exceeded)}
\label{tbl:result1}
\begin{tabular}{c|cccc}
    Number of nodes &5&10&15&20\\
    \hline \hline
    3 phases &0 / 0 / 10&8 / 0 / 2&6 / 0 / 4&2 / 0 / 8\\
    4 phases &0 / 0 / 10&7 / 1 / 2&6 / 0 / 4&7 / 0 / 3\\
    5 phases &1 / 0 / 9&6 / 0 / 4&7 / 0 / 3&5 / 0 / 5\\
    \hline
\end{tabular}
\end{table}

\begin{table}[htbp]
\centering
\caption {Execution times (s)}
\label{tbl:result2}
\begin{tabular}{c|crrrrr}
    Number of nodes & & 5& 10& 15& 20\\
    \hline \hline
            & Mean & -&  3067& 5565& 3699\\
    3 phases& Max  & -& 10480& 8966& 9543\\
            & Min  & -&   258&    1& 3\\
    \hline
            & Mean & -& 3024&  334& 1105\\
    4 phases& Max  & -& 8294& 1103& 2835\\
            & Min  & -& 134&     7&   10\\
    \hline
            & Mean & 32& 4090& 1506& 1266\\
    5 phases& Max  & 32& 9103& 7866& 2586\\
            & Min  & 32&    1&    2& 26\\
    \hline
\end{tabular}
\end{table}

\begin{table}[htbp]
\centering
\caption {Generation times (s) in each phase}
\label{tbl:result3}
\begin{tabular}{c|rrrrr}
    Phase &1&2&3&4&5\\
    \hline
    \hline
    Mean &1.17&2.24&2585.24&188.31&12.49\\
    Max &6.94&24.20&10425.58&1114.09&347.02\\
    Min &0.00&0.00&0.53&3.95&0.03\\
    \hline
\end{tabular}
\end{table}


\section {Conclusion}
\subsection {Summary and Significance of this study}
In this study, we proposed Network Programming for Generalization (NP4G).
We confirmed that NP4G could acquire a bitwise NOT operation program from several training data just by selecting and connecting nodes at random.
NP4G is an inductive inference because it is a method of finding general properties from several examples.
In addition, it has a structure that satisfies the conditions of the structure theorem of "sequence", "selection", and "iteration" for programming, and it is a method that can be expected to automatically acquire arbitrary programs by inductive inference.
Network programming, not limited to genetic methods, demonstrated by NP4G is expected to be widely applied in new fields as a new automatic programming method.
The significance of this study is that it shows the expected realization of versatile and flexible artificial intelligence which can learn by itself from all kinds of knowledge through network programming, and make logical inferences and answer to other problems by making use of that content.

\subsection {Future Issues}
\subsubsection {Turing Completeness}
We explained that NP4G satisfies the conditions of the structure theorem, but in order to prove that NP4G is a method that can obtain arbitrary programs, it is necessary to show that NP4G is Turing complete.
Turing completeness means that a computational model has computational power equivalent to that of a universal Turing machine, that is, it can reproduce arbitrary programs.
If it can be shown to be Turing completeness, it means that we have realized a method that can acquire arbitrary programs by inductive inference.
In addition to a bitwise NOT operation program, it is necessary to actually try whether other programs can be built automatically using NP4G.

\subsubsection {Exploring Search Methods}
In this research, we used accidental search by random generation as a network search method.
Since this method is primitive, the more types and numbers of nodes used, and the more complex the target program, the more time it takes to search.
In the future, by combining with other learning methods such as using reinforcement learning, it is necessary to automatically adjust the types and numbers of nodes suitable for network components and how to combine them in order to improve the efficiency of network construction.

\subsubsection {Evaluation of the Network}
Other learning methods, such as genetic methods and reinforcement learning, have a mechanism to numerically evaluate generated networks and models using evaluation functions.
In this research, it is an evaluation whether or not the target network is generated, and a numerical evaluation by a evaluation function has not yet been performed.
In the future, if we can establish a network evaluation method in NP4G, we can change the construction method by evaluating the network.
In addition, it becomes easy to combine with other learning methods that use evaluation functions.

\subsubsection {Simplification of Generated Networks}
As explained in \ref{sec:sequence}, in a network generated by NP4G, if a node is not derived from the start node, it is not executed.
If there is an algorithm to remove such nodes after generating the network, the generated network can be simplified.

\section*{Acknowledgments}
This research was supported by JST SPRING, Grant Number JPMJSP2130.

\bibliographystyle{unsrt}  
\bibliography{btx_np4g_en}  

\end{document}